\documentclass[aip,cha,preprint]{revtex4-1}
\usepackage{fullpage,amsthm,amsmath,amsfonts,bm,times,color,bbm}
\usepackage{graphicx,subfigure}
\usepackage[english]{babel}

\newcommand{\el}{El Ni\~{n}o}
\newcommand{\la}{La Ni\~{n}a}

\makeatletter

\begin{document}

\title{Percolation framework to describe \el ~conditions
 }
\author{Jun Meng}
\affiliation{Department of Physics, Bar Ilan University, Ramat Gan 52900, Israel}
\affiliation{Solar Energy and Environmental Physics, Blaustein Institutes for Desert Research, Ben-Gurion University of the Negev, Israel}

\author{Jingfang Fan}
\email{j.fang.fan@gmail.com}
\affiliation{Department of Physics, Bar Ilan University, Ramat Gan 52900, Israel}

\author{Yosef Ashkenazy}
\affiliation{Solar Energy and Environmental Physics, Blaustein Institutes for Desert Research, Ben-Gurion University of the Negev, Israel}

\author{Shlomo Havlin}
\affiliation{Department of Physics, Bar Ilan University, Ramat Gan 52900, Israel}

\begin{abstract}
  Complex networks have been used intensively to investigate the flow and dynamics of many natural systems including the climate system. 
  Here, we develop a percolation based measure, the order parameter, to study and quantify climate networks. We find that abrupt transitions of the order parameter  usually occur $\sim$1 year before \el ~ events, suggesting that they can be used as early warning precursors of \el. Using this method we analyze several reanalysis datasets and show the potential for good forecasting of \el. The percolation based order parameter exhibits discontinuous features, indicating possible relation to the first order phase transition mechanism.  
\end{abstract}
\date{\today}

\flushbottom
\maketitle

\thispagestyle{empty}

\textbf{Climate conditions influence the nature of societies and economies. \el~ event,  in particular,
has great influences on climate, which may further cause widespread natural disasters like flood and drought  across the globe. We have just undergone one of the strongest \el~ events since 1948, it brings drought conditions in Venezuela, Australia and more tropical cyclones within the Pacific Ocean. There have been still improvements in the understanding of \el, its climate effects and associated impacts. 
Here, we present a multidisciplinary renaissance combined climate, network and percolation theory to study 
the mechanism of \el. Our method  can forecast 1 year-ahead of \el ~events, with a high prediction accuracy $70 \%$, and a low false alarm. The methodology and results presented here not only facilitate the study of predicting \el~ events but also can bring a fresh perspective to the study of abrupt phase transitions.
}

\section{Introduction}
In the last two decades, complex network became a popular framework to investigate a large variety of real systems, including Internet, social networks, biological networks and financial networks~\cite{watts1998collective,barabasi1999emergence,RevModPhys.74.47,newman2010networks,cohen2010complex}. In recent years, the complex network approach was found to be useful in studying of climate phenomena, using ``climate network''\cite{Tsonis2006,Tsonis2007,yamasaki2008climate,Donges2009a,Donges2009b,Steinhaeuser2010, Steinhaeuser2011,Barreiro2011,Deza2013,ludescher2013very,ludescher2014very,Zemp,Jing2017}. In a climate network, usually, nodes are chosen to be geographic locations and links are constructed based on the similarities between the time variability between pairs of nodes. Climate networks have been used to quantify and analyze the structure and dynamics of the climate system~\cite{Donges2011,gozolchiani2011emergence,Steinhaeuser2012,wang2013dominant,zhou2015teleconnection} . Moreover, climate networks have been used to better understand and even forecast some important climate phenomena, such as monsoon~\cite{Kurth2012,Kurth2013,Kurth2014}, the North Atlantic Oscillation~\cite{Guez2012,Guez2013} and \el ~\cite{tsonis2008topology,yamasaki2008climate,gozolchiani2011emergence, Radebach,ludescher2013very,ludescher2014very}.

\el ~is probably the most influential climate phenomenon on interannual time scales~\cite{dijkstra2005,clarke2008,sarachik2010,wangc2012}. During \el, the eastern Pacific ocean is getting warmer by several degrees, impacting the local and global climate. \la~ is cold anomaly over the \el ~region. The \el ~activity is quantified, for example, by the Oceanic Ni\~{n}o Index (ONI), which is NOAA's primary indicator for monitoring \el ~and \la. \el ~ can trigger many disruptions around the globe and in this way affect various aspects of human life. These include unusual weather conditions, droughts, floods, declines in fisheries, famine, plagues, political and social unrest, and economic changes. Global impacts of \el ~had been investigated in by Halpert et al ~\cite{Halpert}. However, the mechanism through which \el ~influences the global climate and the impact of \el ~are still not fully understood. Here we propose a percolation framework analysis to describe the structure of the global climate system during \el, based on climate networks.

Percolation theory is also used to analyze the behavior of connected clusters in a network \cite{stauffer1994introduction,cohen2010complex,bunde2012fractals}. The applications of percolation theory covers many areas, such as, optimal path, directed polymers, epidemics, immunization, oil recovery, and nanomagnets. In the framework of percolation theory one may define phase transition based on simplest pure geometrical considerations.

In the present study, we construct a sequence of monthly-shifting-climate networks by adding links one by one according to the similarities between nodes. More specifically, the nodes which are more similar (based on their temperature variations) will be connected first. We statistically found that around one year prior to the onset of \el, the climate network undergoes a first order phase transition (i.e., exhibiting a significant discontinuity in the order parameter), indicating that links with higher similarities tend to localize into two large clusters, in the higher latitudes of the northern and southern hemispheres. However, during \el ~times, there is only one big cluster via tropical links. We find that indications the discontinuity in the order parameter is closely related to the ONI.

\section{Climate Network}
Our analysis is based on the daily near surface ($1000$ hPa) air temperature of ERA-Interim reanalysis~\cite{dee2011era}. We pick 726 grid points that approximately homogeneously cover the entire globe~\cite{gozolchiani2011emergence}; these grid points are chosen to be the nodes of our climate network. For each node (i.e., longitude-latitude grid point), daily values within the period 1979 - 2016 are used, from which we subtract the mean seasonal cycle and divide by the seasonal standard deviation. Specifically, given a record $\tilde{T}^{y}(d)$, where $y$ is the year and $d$ stands the day (from 1 to 365), the filtered record is defined as,
\begin{equation}
T^{y}(d) =  \frac{\tilde{T}^{y}(d) - mean(\tilde{T}(d))}{std(\tilde{T}(d))},
\label{eq1}
\end{equation}
where  ``mean'' and ``std'' are the mean and standard deviation of the temperature on day $d$ over all years.

To obtain the time evolution of the strengths of the links between each pair of nodes, we define, the time-delayed cross-correlation function as,
\begin{equation}
C_{i,j}(-\tau) =  \frac{<T_{i}(d)T_{j}(d-\tau)> -<T_{i}(d)><T_{j}(d-\tau)>}{\sqrt{(T_{i}(d) - <T_{i}(d)>)^2} \cdot \sqrt{(<T_{j}(d-\tau) - <<T_{j}(d-\tau)>)^2}},
\label{eq2}
\end{equation}
and
\begin{equation}
C_{i,j}( \tau) =  \frac{<T_{i}(d-\tau)T_{j}(d)> -<T_{i}(d-\tau)><T_{j}(d)>}{\sqrt{(T_{i}(d-\tau) - <T_{i}(d-\tau)>)^2} \cdot \sqrt{(<T_{j}(d) - <<T_{j}(d)>)^2}},
\label{eq3}
\end{equation}
where $\tau$ is the time lag between 0 and 200 days. Note, that for estimating the cross-correlation function at day $d$, only temperature data points prior to this day are considered. We then define the link's weight as the maximum of the cross-correlation function $max(C_{i,j}(\tau))$. 

\section{Percolation}
In lattices model, a percolation phase transition occurs if the systems' dimension is larger than one~\cite{bunde2012fractals}. The system is considered percolating if there is a path from one side of the lattice to the other, passing through occupied bonds (bond percolation) or sites (site percolation). The percolation threshold usually depends on the type and dimensionality of the lattice. However, for the network system no notion of side exists. For this reason, a judgment condition to verify whether the system is percolating is the existence of a giant component (cluster) containing $O(N)$ nodes, where $N$ is the total number nodes in the network. If two nodes are in the same cluster then there is at least one path passing through them.

In this section, we discuss the construction of the climate networks, and study the evolution of clusters. Initially, given $N=726$ isolate nodes, links are added one by one according to the link strength, i.e., we first add the link with the highest weight, and continue selecting edges ordered by decreasing weight. During the evolution of our network, we measure the size of the normalized largest cluster $s_{1} =S_{1}/N$ and the susceptibility $\chi$, where $S_1$ represents the size of the largest component. The susceptibility of the climate network (the average size of the finite clusters) is defined as ~\cite{stauffer1994introduction},
\begin{equation}
\chi = \frac{\sum_{s}^{'} s^{2} n_{s}(C)}{\sum_{s}^{'} s n_{s}(C)},
\label{sus}
\end{equation}
where $n_{s}(C)$ denotes the average number of clusters of size $s$ at edge's weight $C$, and the prime on the sums indicates the exclusion of the largest cluster in each measurement.

Since our network is finite, we use the following procedure to find the percolation threshold. We first calculate, during the growth process, the largest size change of the largest cluster:
\begin{equation}
\Delta \equiv \frac{1}{N}\max\left[S_1(2)-S_1(1),\cdots,S_1(T+1)-S_1(T),\cdots\right].
\label{eq5}
\end{equation}
The step with the largest jump is defined as $T_c$. The percolation transition in the network is characterized by $\Delta$ and $T_c$ corresponds to its transition point.

\section{Results}

\begin{figure}
\begin{centering}
\includegraphics[width=1.0\linewidth]{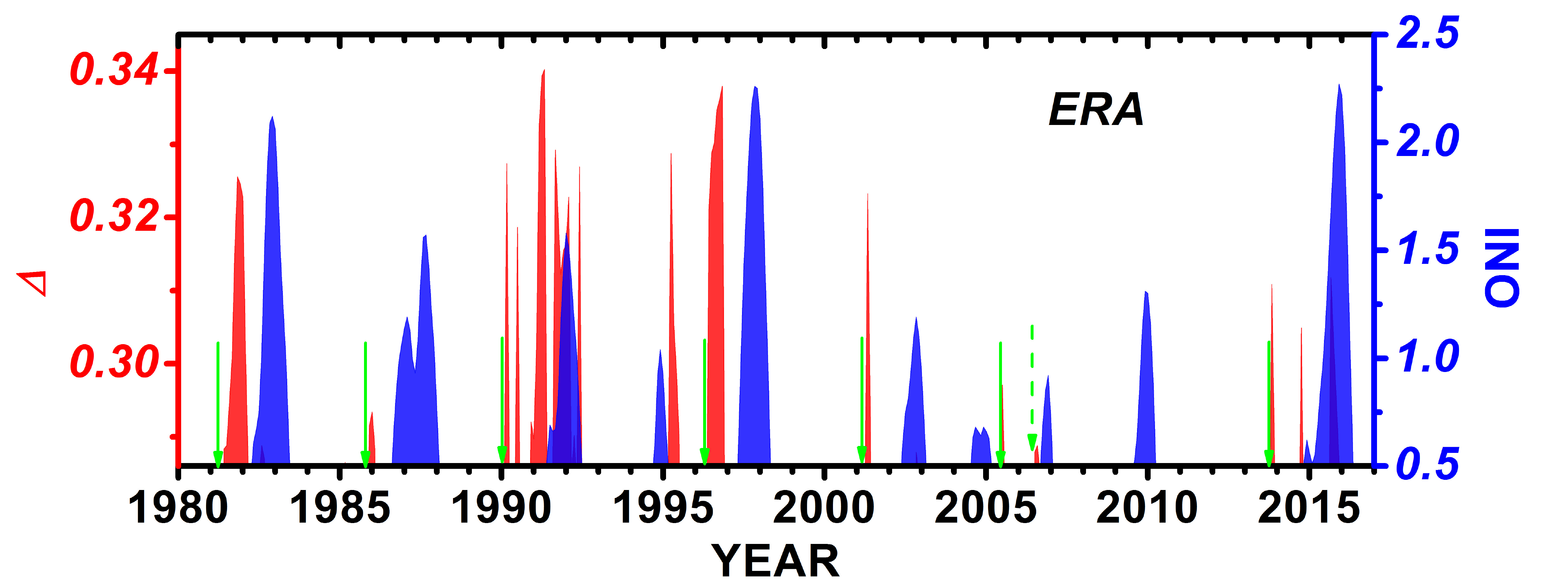}
\caption{\label{Fig:1}(Color online). The percolation forecasting  scheme and power based on the near surface temperature of the ERA-Interim dataset \cite{dee2011era}.  We compare the largest gap of the largest cluster $\Delta$ during the climate network evolution  with a threshold $\Theta = 0.286$ (red curve, left scale) and the ONI (blue curve, right scale) between January 1980 and September 2016. When the $\Delta$ is above the  threshold, $\Theta$, we give an alarm and predict that an \el~ event  will start in the following calendar year.  Correct predictions are marked by green arrows and false alarms by dashed arrows.}
\end{centering}
\end{figure}

 \begin{figure}
\begin{centering}
\includegraphics[width=1.0\linewidth]{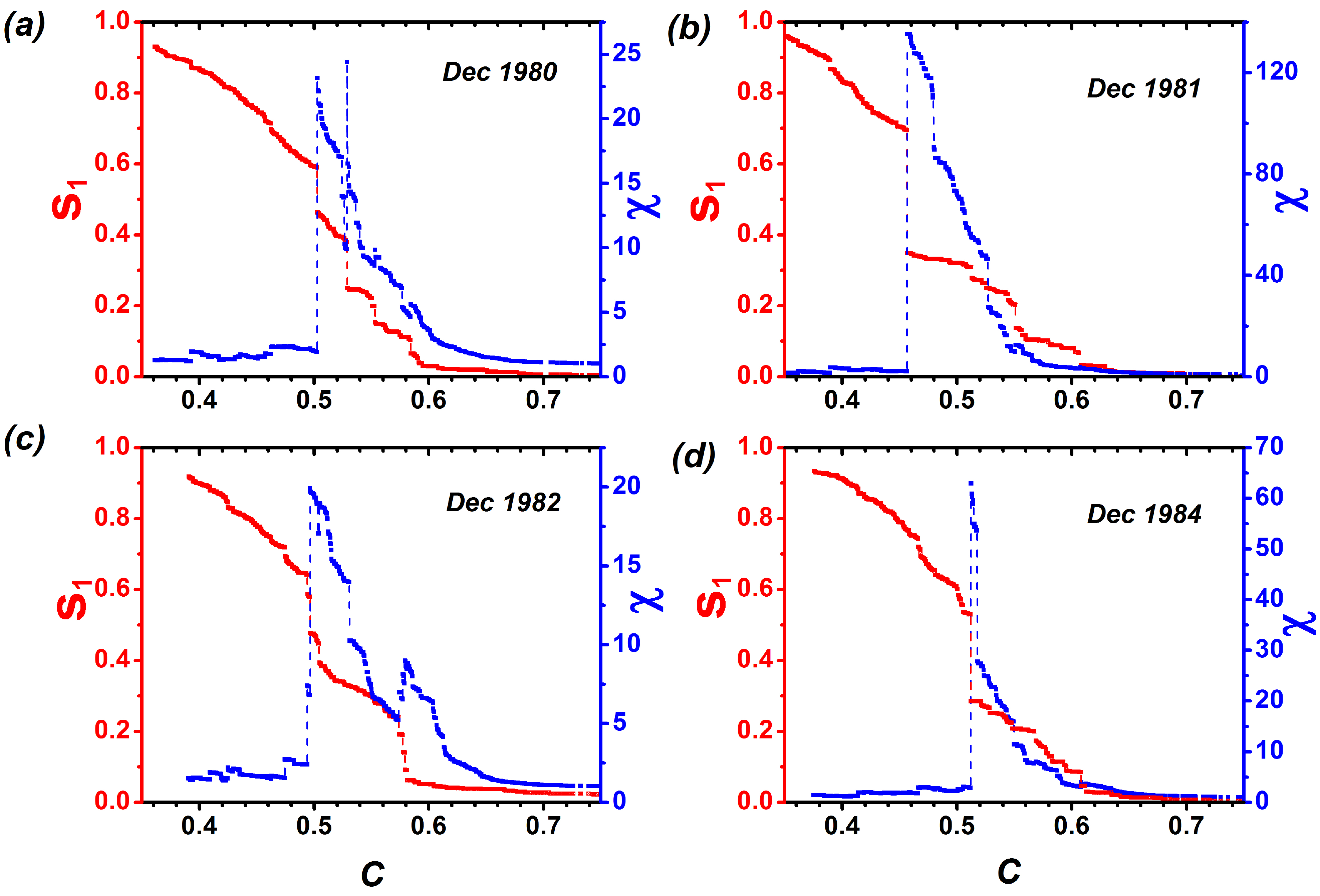}
\caption{\label{Fig:2}(Color online). The largest cluster $s_1$ (red curve, left scale) and the susceptibility $\chi$ (blue curve, right scale) as a function of the link strength $C$. (a) For the network two years before \el~ episode, where the end time is Dec 1980; (b) for the network one year before \el~ episode, where the end time is Dec 1981; (c) for the network during  \el~ episode, where the end time is Dec 1982; (d) for the network one year after \el~ episode, where the end time is Dec 1984. Note the largest jump in $\Delta$ one year prior to \el~ event.}
\end{centering}
\end{figure}

For each network, we obtain $\Delta$, and find that, usually around one year ahead of the beginning of \el, the climate network has the largest $\Delta$. This feature is used here for forecasting the inception of an \el~ event in the following year. To this end, we place a varying horizontal threshold $\Delta=\Theta$ and mark an alarm when $\Delta$ is above threshold, outside an \el~ episode. Fig.~\ref{Fig:1} demonstrates the forecasting power where the red curve depicts $\Delta$, and the blue curve is the ONI; correct predictions are marked by green arrows. The lead time between the prediction and the beginning of the \el~ episodes is $1.05 \pm 0.18 $ year. Our method forecasts $7$ out of $10$ events.
Note the similarity in the power of forecasting to that of Ludescher et al~\cite{ludescher2014very}.

Next, we concentrate on specific \el~ events to illustrate the evolving cluster structure through ~\el. We first focus on one of the strongest \el~ event, the $1982-1983$ event. In Fig.~\ref{Fig:2} we show for this event, $s_1$ and $\chi$ as a function of link strength $C$ two years and one year before ~\el, during \el~ and one year after \el. We find that $s_1$ exhibits the largest jump in $\Delta$ about one year before \el; the jump in $\chi$ also becomes very large at the same point. The two quantities yields the same percolation threshold, strengthening the confidence of the threshold value. 

\begin{figure}
\begin{centering}
\includegraphics[width=1.0\linewidth]{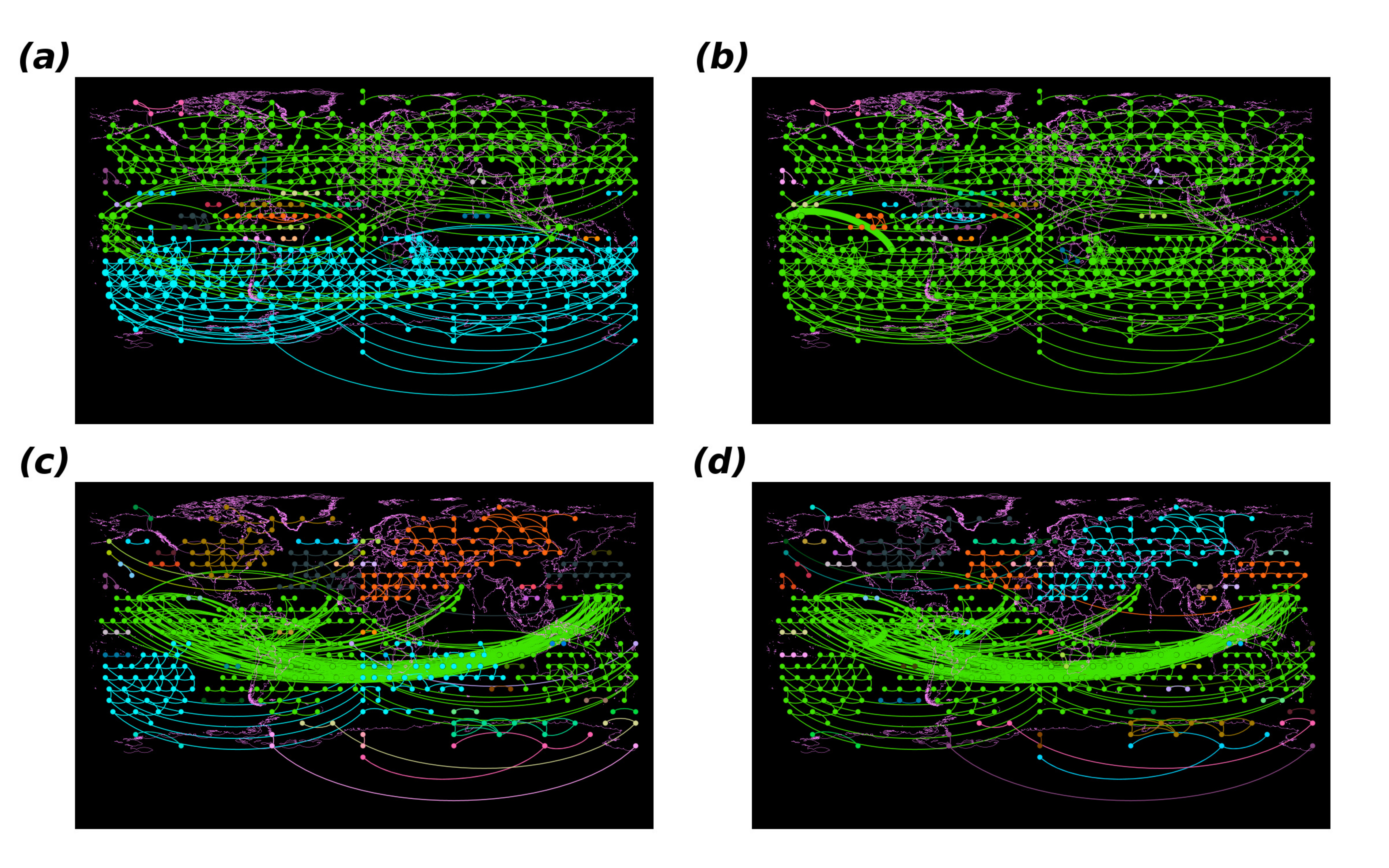}
\caption{\label{Fig:2a}(Color online). The cluster structure on map at the percolation threshold for the network one year before the \el~ event (Dec 1981). (a) Before the critical link was added; (b) after the critical link (marked by thicker green line) was added. (c) and (d) are the cluster structure at the percolation threshold for the network during \el~ episode (Dec 1982). Different colors represent different clusters, especially, the green represents the largest cluster and the blue represents the second largest cluster. }
\end{centering}
\end{figure}

Fig.~\ref{Fig:2a} (a) shows the climate network cluster structure in the globe map at the percolation threshold one year before \el~ event. It seems like the equatorial region separates the network into two communities, Northern and Southern hemispheres, where the nodes with green color indicate the largest cluster and the blue indicates the second largest cluster; after the critical link adding (marked by thicker green line), the largest and second largest cluster merge, and the new largest cluster approximately covers the entire globe (Fig.~\ref{Fig:2a} (b)). We find that typically during the \el~ event (Fig.~\ref{Fig:2}(c)) $s_1$ does not exhibit a large jump at the percolation threshold. Fig.~\ref{Fig:2a} (c) shows the cluster structure at the percolation threshold. There are more edges in the tropical zone. This is since during the \el ~period, the nodes in low latitudes are drastically affected by the \el, resulting in higher cross-correlation. Therefore, we do not find a large gap in the percolation of the network.

\begin{figure}
\begin{centering}
\includegraphics[width=1.0\linewidth]{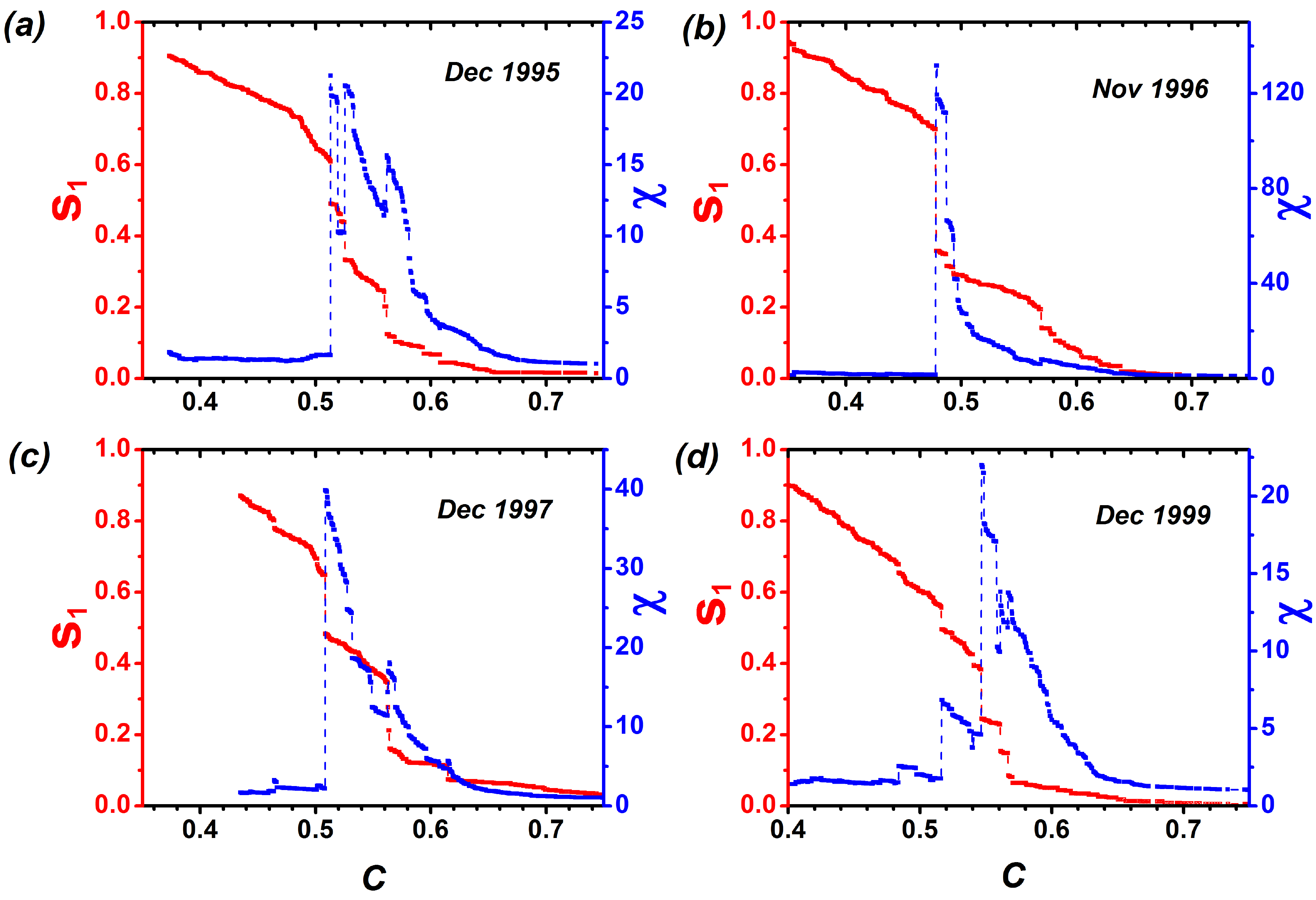}
\caption{ \label{Fig:2e}(Color online). Same as Fig.~\ref{Fig:2} for the strong 1997-1998 \el~ event. }
\end{centering}
\end{figure}

Another example for the evolution of the network (represented by $s_1$ and $\chi$ vs. $C$) during the $1997-1998$ \el~ event is shown in Fig.~\ref{Fig:2e}. Also here we find that one year before event there is a large gap in $s_1$ (Fig.~\ref{Fig:2e} (b)), however, during the \el~event, two years before the event and one year after the event, the gap becomes smaller (Fig.~\ref{Fig:2e} (a)(c)(d)).

%

\begin{figure}
\begin{centering}
\includegraphics[width=0.75\linewidth]{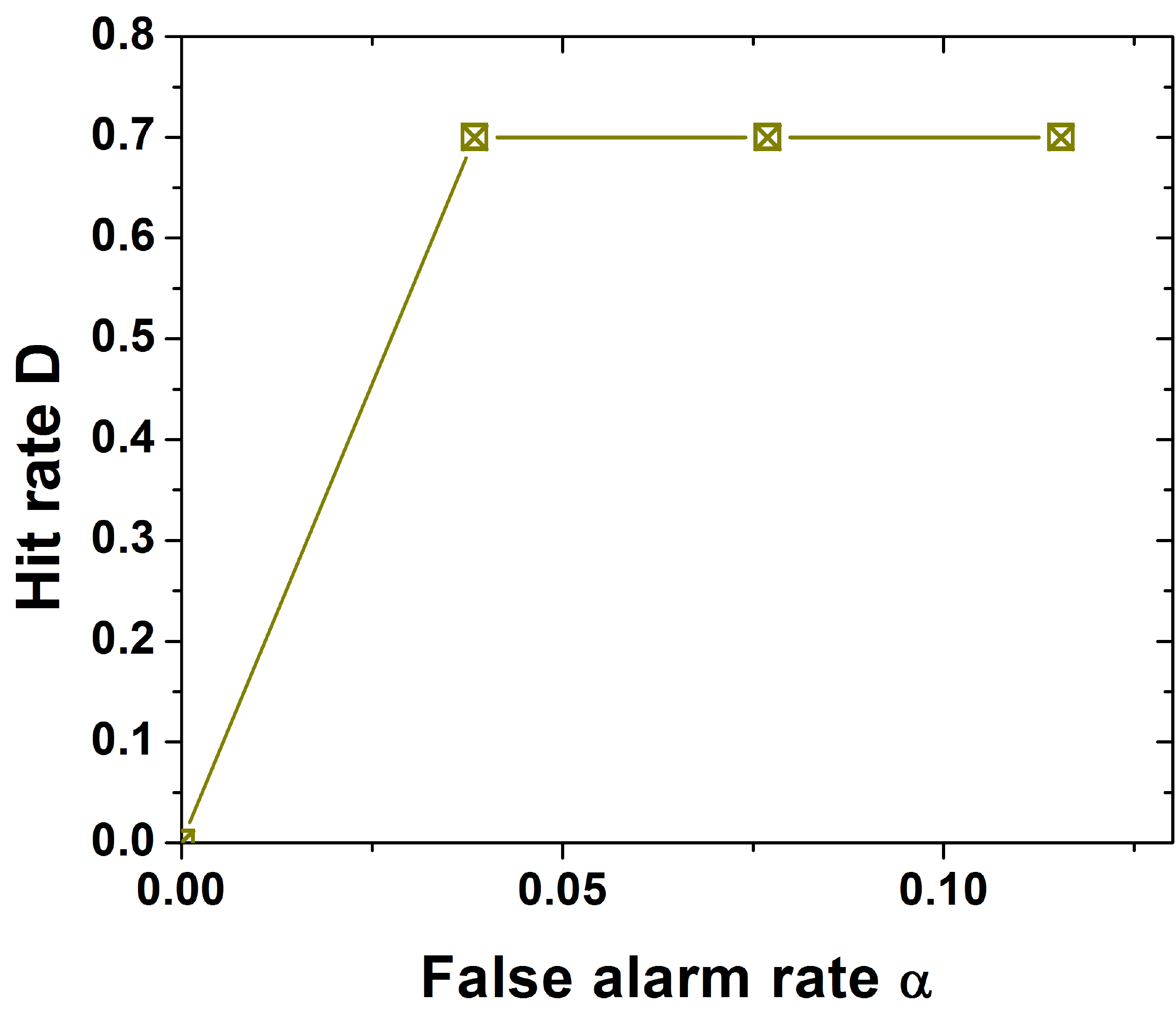}
\caption{\label{Fig:3}(Color online). The prediction accuracy for out method. For the four lowest false-alarm rates $\alpha$ = $0, 1/27, 2/27, 3/27$ the best hit rates $D$.  }
\end{centering}
\end{figure}

Following the above, we assume that large $\Delta$ is an alarm forecasts that  \el~ will develop in the following calendar year. In the case of multiple alarms in the same calendar year, only the first one is considered. The alarm results in a correct prediction, if in the following calendar year an \el~ episode actually occurs; otherwise it is regarded as a false alarm. There were 10 \el~ events (years) between 1979 and 2016 and additional 27 non-\el~ years. To quantify the accuracy for our prediction, we use the Receiver Operating Characteristic (ROC)-type analysis~\cite{ludescher2013very} when altering the magnitude of the threshold and hence the hit and false-alarm rates. Fig.~\ref{Fig:3} shows, the best hit rates for the false-alarm rates 0, 1/27, 2/27, and 3/27. The best performances are for (i) thresholds $\Theta$ in the interval between 0.286 and 0.289, where the false-alarm rate is 1/27 and the hit rate is 0.7, for (ii) thresholds between 0.264 and 0.266, where the false-alarm rate is 2/20 and the hit rate is 0.7, and (iii) for thresholds between 0.223 and 0.26, where the false-alarm rate is 3/20 and the hit rate is 0.7.

To further test our results we also applied the same method for different datasets, the NCEP/NCAR reanalysis dataset ~\cite{kalnay1996ncep}, and the JRA-55 dataset~\cite{Onogi2007}. To allow simple comparison to the ERA-Interim reanalysis, we only consider the last 37 years (1980-2016). The prediction accuracy are summarized in Table. ~\ref{comparison}. We basically find very similar results for all three different reanalysis datasets, strengthening the confidence in our prediction method. 

\begin{table}[htbp]
\centering  %
\caption{\label{comparison} The forecast accuracy for different reanalysis datasets, based on the Receiver Operating Characteristic (ROC)-type.}
\begin{tabular}{  c | c | c  }
\hline
Dataset & Hit rates $D$ & False-alarm rates $\alpha$ \\ \hline
ERA-Interim & 0.7 & 1/27  \\ 
NCEP/NCAR reanalysis & 0.6 & 1/27  \\ 
JRA-55 & 0.6 & 1/27  \\ \hline
\end{tabular}
\end{table}

\begin{figure}
\begin{centering}
\includegraphics[width=1.0\linewidth]{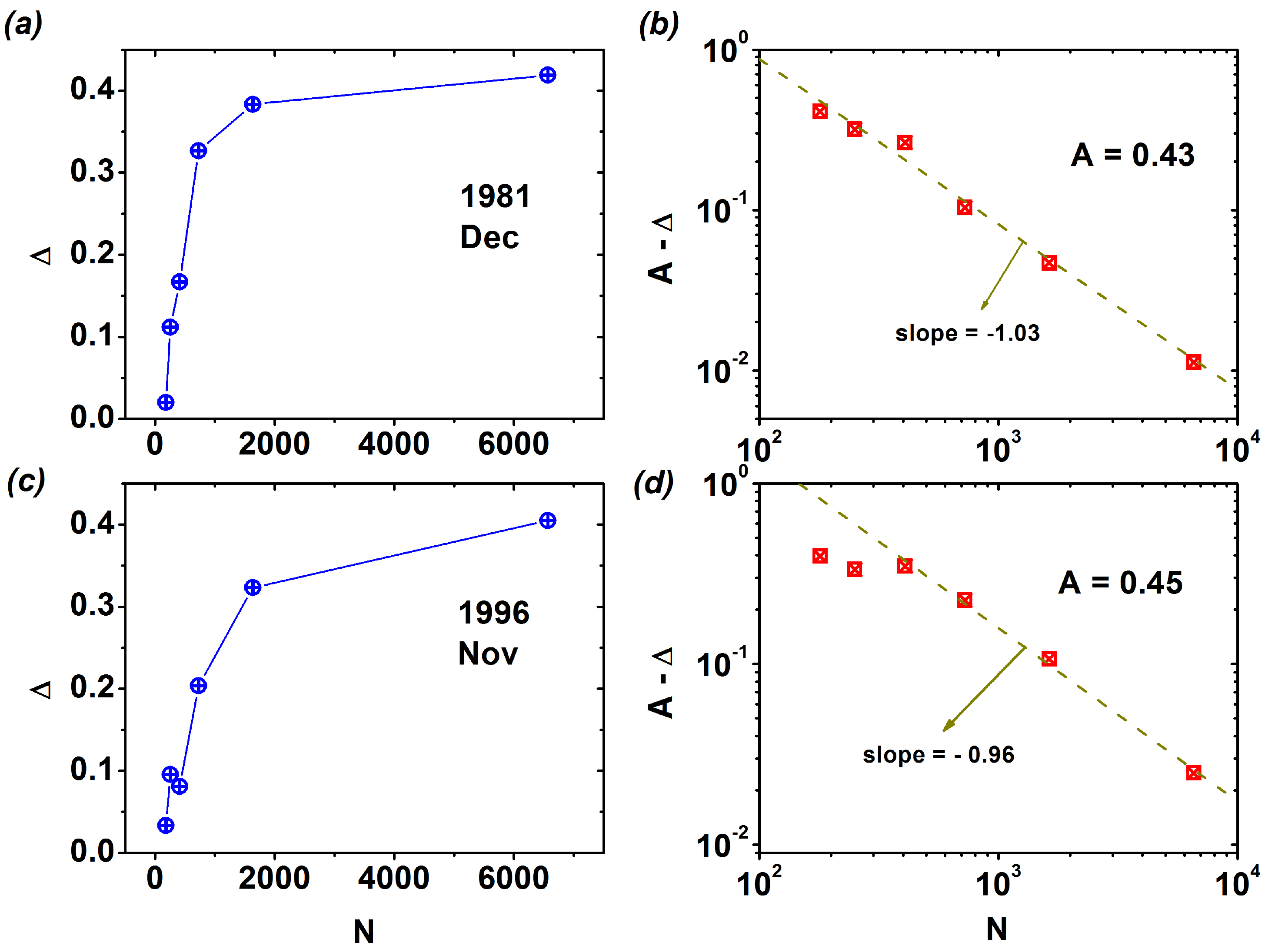}
\caption{\label{Fig:4}(Color online). The finite size effects for the networks before \el~ episode. The largest gap $\Delta(N)$ as a function of system size $N$ for (a) Dec. 1981 and (b) Nov 1996.  (b),(d) Log-log plot of $A-\Delta(N)$ versus $N$, indicating possible scaling law with scaling exponent $\sim$1; see Eq.~(\ref{eq6}).} 
\end{centering}
\end{figure}

To further test the order of the percolation phase transitions in the climate network before \el ~event, we study the finite size effects of our network, and suggest that the transition is a first order phase transition. We change the system's size by altering the resolution of nodes, and at the same time make sure that every node in a given covers the same area on the global (i.e., we are fewer nodes at the high latitudes). First, we define the resolution (in degree latitude) at the Equator as $r_0$ and then find that the number of nodes is $n_0 = 360/r_0$. Then the number of nodes in latitude $r_0m$ is $n_m = n_0cos(r_0m)$, where $m\in [-90/r_0, 90/r_0]$. The total number of nodes is then $N = \sum\limits_{m=0}^{m=90/r_{0}} 2n_m - n_{0}$. We choose $r_0$ to be $(15, 12.5, 10, 7.5, 5, 2.5)$ $^\circ$, which yields $N=(180, 251, 408, 726, 1634, 6570)$. We then calculate $\Delta$ as a function of the system size $N$. If $\Delta$ approaches zero as $N\rightarrow\infty$, the corresponding giant component is assumed to undergo a continuous percolation; otherwise, the corresponding percolation is assumed to be discontinuous. This is since it suggests that the order parameter $s_1$ has a non-zero discontinuous jump at the percolation threshold \cite{Nagler2011,fan2014general}. The results of $\Delta$ as a function of the system size $N$ are shown in Fig.~\ref{Fig:4} for two  \el~ events considered above. The results suggest a discontinuous percolation since $\Delta(N)$ tends to a non-zero constant (Fig.~\ref{Fig:4} (a) and (c)). We also find that $\Delta$ follows a scaling form,
\begin{equation}
A - \Delta(N) \sim  N^{-\beta}.
\label{eq6}
\end{equation}
where $A$ is a constant and $\beta$ is a critical exponent. Fig.~\ref{Fig:4} (b) and (d) show the related results where we find that $\beta$ is very close to $1$, implying it might be an universal scaling exponent.

It has been pointed out that a random network always undergoes a continuous percolation phase transition during a random process \cite{bollobas2001random}. The question whether percolation transitions could be discontinuous has attracted much attention. Discontinuous percolation in networks was reported in the framework of the explosive percolation model ~\cite{achlioptas2009explosive}. However, later studies questioned this finding~\cite{PhysRevLett.105.255701,Riordan2011,radicchi2010explosive,grassberger2011explosive,araujo2011tricritical,chen2011explosive,nagler2012continuous,PhysRevE.85.061110}. Interestingly, our results indicate the possibility of first order phase transition in climate networks.

\section*{Conclusions }

To summarize, a time-evolving weighted climate network is constructed based on near surface air temperature time series. A percolation framework to study the cluster structure properties of the climate network is put forward. We find that the structure of the network changes violently approximately one year ahead of \el~ events---we suggest to use such abrupt transitions to forecast  \el~ events. The percolation description of climate system (as reflected by the surface air temperature records) highlight the importance of such network techniques to understand and forecast \el~ events. Based on finite size scaling analysis, we also find that the percolation process is discontinuous. The methodology and results presented here not only facilitate the study of predicting \el ~events but also can bring a fresh perspective to the study of abrupt  phase transitions.

\begin{acknowledgments}
J. Fan thanks the fellowship program funded by the Planning and Budgeting Committee of the Council for Higher Education of Israel. We acknowledge the
MULTIPLEX (No. 317532) EU project, the Israel Science Foundation, ONR and DTRA for financial support.
\end{acknowledgments}

\end{document}